\documentclass[reprint,aip,pop,graphicx,numerical,amsfonts,amsmath,amssymb,onecolumn]{revtex4-1}

\usepackage{bm}
\usepackage[dvips]{graphicx}
\usepackage{lineno}
\usepackage{color}
\usepackage{soul}
\usepackage{booktabs}
\usepackage{multirow}
\sethlcolor{yellow}

\newcommand{\biota}{\iota
                     \hskip-.15ex{\hbox to 0pt{\hss {\leavevmode
                     \hbox{\raise -.60ex \hbox{{\tt \'{}}}}}}}
                     \hskip.37ex{\hbox to 0pt{\hss {\leavevmode
                     \hbox{\raise -.50ex \hbox{{\tt \'{}}}}}}}}
\newcommand{\delop}[2]{\frac{\partial #1}{\partial #2}}
\newcommand{\ve}{\bm{v}_{\rm E}}
\newcommand{\vb}{v_{\rm B}}
\newcommand{\vbh}{\hat{\bm{v}}_{\rm B}}
\newcommand{\exb}{\bm{E}\times\bm{B}}
\newcommand{\Er}{E_{\rm r}}
\newcommand{\zlocal}{\bm{z}^{(4)}}
\newcommand{\dzlo}{\dot{\bm{z}}^{(4)}}

\begin{document}

\preprint{00}

\title{Effects of magnetic drift tangential to magnetic surfaces
  on neoclassical transport in non-axisymmetric plasmas}



\author{Seikichi Matsuoka}
\email[]{matsuoka@rist.or.jp}
\affiliation{Research Organization for Information Science and Technology, 6F Kimec-Center Build., 1-5-2 Minatojima-minamimachi, Chuo-ku, Kobe, 650-0047 Japan}

\author{Shinsuke Satake}
\author{Ryutaro Kanno}
\affiliation{National Institute for Fusion Science, 322-6 Oroshi-cho, Toki, 509-5292 Japan}
\affiliation{Department of Fusion Science, SOKENDAI (The Graduate University for Advanced Studies), 322-6 Oroshi-cho, Toki, 509-5292 Japan}

\author{Hideo Sugama}
\affiliation{National Institute for Fusion Science, 322-6 Oroshi-cho, Toki, 509-5292 Japan}


\date{\today}

\begin{abstract}
In evaluating neoclassical transport by radially-local simulations,
the magnetic drift tangential to a flux surface is usually ignored
in order to keep the phase-space volume conservation.
In this paper, effect of the tangential magnetic drift on the local
neoclassical transport are investigated.
To retain the effect of the tangential magnetic drift in the local treatment of neoclassical transport,
a new local formulation for the drift kinetic simulation is developed.
The compressibility of the phase-space volume caused by the tangential magnetic drift
is regarded as a source term for the drift kinetic equation,
which is solved by using a two-weight $\delta f$ Monte Carlo method
for non-Hamiltonian system
[G.~Hu and J.~A.~Krommes, Phys. Plasmas \textbf{1}, 863 (1994)].
It is demonstrated that the effect of the drift is negligible for
the neoclassical transport in tokamaks.
In non-axisymmetric systems, however,
the tangential magnetic drift substantially changes the dependence
of the neoclassical transport on the radial electric field $\Er$.
The peaked behavior of the neoclassical radial fluxes around $\Er = 0$
observed in conventional local neoclassical transport simulations is removed by taking the tangential magnetic drift into account.
\end{abstract}


\maketitle 



\section{Introduction}
\label{sec:intro}
Neoclassical transport caused by Coulomb collisions in torus plasma
is fundamental for a magnetically
confined plasma since it determines an irreducible minimum for the plasma transport.
It also plays a key role in determining the radial electric field
through the ambipolar condition of the neoclassical particle flux
when non-axisymmetric devices such as stellarators and heliotrons are considered.
In addition, the neoclassical viscosity caused by non-uniform magnetic field influences plasma parallel flows.

The neoclassical transport theory is based on the drift kinetic equation,
in which the fast gyration of the plasma particle is removed.
Many analytic and numerical evaluations have been done for axisymmetric tokamaks
and non-axisymmetric devices.~\cite{Balescu1988,Hirshman1981,Hinton1976,
Shaing1983,Sugama1996,Sugama2002,Hirshman1986,CDBeidlerandWDD'haeseleer1995}
For this purpose,
additional assumptions are usually made in the drift kinetic equation.
At first, the higher order radial drift is neglected.
This enables ones to solve the ``radially local'' drift kinetic equation,
leading to ``local'' neoclassical transport,
where ``local'' means that the drift kinetic equation and the neoclassical transport
is only determined by its radially local parameters.
Second,
the tangential component of the magnetic drift:
\begin{equation}
\label{eq:vbh}
\vbh \equiv {\bm v}_{\rm B} - \left({\bm v}_{\rm B}\cdot\nabla s\right) {\bm e}_{s}
\end{equation}
is omitted, where ${\bm v}_{\rm B}$ is the magnetic drift composed
of the $\nabla B$ drift and the curvature drift,
$s$ is a label of magnetic flux surfaces,
and $\bm{e}_s$ is the covariant basis vector in $s$ direction.
Third, the mono-energetic particle assumption is of importance.
This assumes that the particle velocity $v$, or kinetic energy $mv^2/2$ is unchanged
along the particle orbit.
Finally, $\exb$ drift is assumed to be incompressible to conserve the phase-space volume.
With these assumptions,
the evaluation of the neoclassical transport becomes much easier
since the drift kinetic equation described in five-dimensional
phase space is reduced to that in three-dimensional phase space.



As mentioned above,
the neoclassical transport simulations are based on many assumptions,
which are interdependent.
The main purpose of this paper is to reconsider the validity of the approximations,
especially with respect to the tangential magnetic drift, $\vbh$.
Depending on the approximations made in the drift kinetic equation,
various kinds of the drift kinetic equation and the particle orbit appear
in this paper.
(A) The drift kinetic equation without all the assumptions described above.
Since the equation includes the higher order radial drift term in this case,
the neoclassical transport with the finite orbit width (FOW) effect can be evaluated.~\cite{Satake2008}
The neoclassical transport is also called a {\it global} one
due to the fact that it involves the radially global effect in it.
It should be noticed that
since $\dot{v}$ is in proportion to the product of the radial drift and the radial electric field $\Er$,
we also call $\dot{v}$ the FOW effect in this paper.
(B) The drift kinetic equation without the radial drift term.
Since the radial drift term is neglected,
the drift kinetic equation becomes local, and the local neoclassical transport is obtained.
We refer to this particle orbit as zero orbit width (ZOW) orbit
in order to distinguish it from other kinds of the local orbit.
(C) The drift kinetic equation in ZOW limit with $\vbh = 0$.
The particle orbit and the drift kinetic equation in this limit
have many preferable features, as described later in Sec.~\ref{sec:dke},
some authors evaluate this type of the local neoclassical transport.
We would like to call it the zero magnetic drift (ZMD) limit.
(D) ZMD limit with the mono-energetic particles and incompressible $\exb$ drift.
In this limit, the particle orbit reduces to the same one that is adopted
in a widely-used neoclassical transport code, DKES.~\cite{Hirshman1986,VanRij1989}
We call this particle orbit as the DKES-like orbit.


Conventionally, the neoclassical transport has been
evaluated locally, and DKES-like orbit has been adopted in many codes.
This is justified in a typical torus plasma
if the radial drift term is negligibly small.
Recently, however, several authors have pointed out that
there are some cases where the approximations in the conventional local neoclassical transport models are violated.
For example, the FOW effect becomes significant
near the axis of a tokamak due to the potato orbit.~\cite{Satake2002}
The electron FOW affects the neoclassical transport in a high electron temperature stellarator
due to its complicated orbit and low collisionality.~\cite{Matsuoka2011a}
Also, the mono-energy assumption may cause underestimation of the fraction of the helically-trapped particles
in a quasi-symmetric stellarator when the radial electric field is finite.~\cite{Landreman2011}

So far, efforts have been made
to investigate the effects of the FOW and/or the mono-energetic particle,
while the effect of the tangential magnetic field on the local neoclassical transport
has not been considered.~\cite{Matsuoka2011a,Landreman2013,Landreman2014}
Although conventional local neoclassical
codes provide reliable results in many cases,
there seems to be several situations when the magnetic drift needs to be included,
{\it e.g.,} a resonant behavior of the magnetic drift with $\exb$ drift.
The resonant behavior called the poloidal resonance
between the tangential magnetic drift and $\exb$ drift occurs in a non-axisymmetric magnetic field configuration.
Dependence of the neoclassical transport on the radial electric field is qualitatively
varied by the poloidal resonance.
However, the influence of the tangential magnetic drift on the neoclassical transport is not fully clarified
since there are no local neoclassical transport models which include the effect.
This makes it difficult to compare neoclassical transport models with and without the tangential magnetic drift.
For example, when comparing the global neoclassical transport to the local one in DKES-like limit,
the compressible $\exb$ drift, finite $\dot{v}$ and $\vbh$ in addition to the effect of the radial motion
simultaneously affect the neoclassical transport of the global model.
In other words, there exists a large gap between the global neoclassical transport model,
in which the effect of the tangential magnetic drift is included,
and conventional local models ignoring the effect.
It is necessary to bridge the gap by developing a local neoclassical transport model based on the drift kinetic equation in the ZOW limit in order to explore the effect.

In this paper,
we present a new formulation of the local drift kinetic simulation in the ZOW limit,
where the tangential magnetic drift term is retained while the radial drift term is ignored.
Due to the tangential magnetic drift, the phase-space volume, and thus the particle number are not conserved.
The resultant local drift kinetic equation becomes non-Hamiltonian,
and the compressibility of the phase-space volume acts as a source term.
Hu and Krommes prescribes the two-weight $\delta f$ method appropriate
for such non-Hamiltonian system with an arbitrary source/sink term.~\cite{Hu1994}
Based on their work,
we develop a numerical code for the local neoclassical transport with $\vbh$.
The code developed here requires less computational cost than global ones with the FOW effect
due to the local feature of the code.
It provides a more accurate method to evaluate the neoclassical transport in plasmas
where the tangential magnetic drift becomes significant.
Another advantage of the code is that
the particle orbit in the code can be switched among the ZOW, ZMD and DKES-like limit models.
This enables us to investigate the effect of the particle orbit on the local neoclassical transport.
We investigate the neoclassical transport in an axisymmetric and a non-axisymmetric plasmas
using the code.
It is found that the neoclassical transport in the ZOW limit
is almost the same as that in DKES-like limit in an axisymmetric case.
This suggests that $\vbh$ is not significant in axisymmetric tokamak as expected.
On the other hand, the local neoclassical transport in ZOW limit is demonstrated
to show a radial electric field dependence, or the poloidal resonance at a finite $\Er$,
which has not been seen in the local neoclassical transport.
When using the ZMD and/or DKES-like limit orbits,
a large peak of the neoclassical radial flux is observed at $\Er = 0$.
Such a large neoclassical transport is removed in the ZOW limit
due to the effects of the tangential magnetic drift.
As a result, the neoclassical transport in the ZOW limit approaches to that in the global FOW model.

The remaining part of this paper is organized as follows.
The drift kinetic equations with FOW effect and in various local limits
(ZOW, ZMD, and DKES-like limits)
are described in Sec.~\ref{sec:dke}.
The property of the phase-space conservation in each limit
is also presented.
The two-weight $\delta f$ Monte Carlo method for non-Hamiltonian system
is described in Sec.~\ref{sec:method}.
Numerical results for axisymmetric case and non-axisymmetric case
are presented in Sec.~\ref{sec:axisym} and ~\ref{sec:assym}.
A summary is given in Sec.~\ref{sec:summary}.

\section{Drift kinetic equation for neoclassical transport}
\label{sec:dke}

\begin{table*}[t]
 \centering
 \caption{\label{tb:dke}
   Comparison of guiding-center orbits and conservation properties included in the global and local drift kinetic models.
   Model equations for models (A), (B), (C) and (D) are described in corresponding subsections of Sec.~\ref{sec:dke}.
   In the table, {\it Comp.} and {\it Incomp} denote {\it compressible} and
   {\it incompressible}, respectively.}
 \begin{tabular}{l||@{\hspace{0.3cm}}l|@{\hspace{0.4cm}}l|@{\hspace{0.4cm}}l|@{\hspace{0.3cm}}l|@{\hspace{0.4cm}}l}
\hline
& Global & \multicolumn{4}{c}{Local} \\[4pt] \hline\hline
Model & (A) FOW & (B) ZOW & (C) ZMD & ZMD & (D) DKES-like \\[4pt]
& {\footnotesize (Finite Orbit Width)} & {\footnotesize (Zero Orbit Width)} & {\footnotesize (Zero Magnetic Drift)} & + mono-energy & \\[2pt] \hline
\multirow{3}{20mm}{Particle orbit}
& \multirow{3}{20mm}{Full orbit}
& \multirow{3}{30mm}{finite $\vbh$}
& \multirow{3}{30mm}{$\vbh = 0$}
& \multirow{3}{25mm}{$\vbh = 0$,\\[2pt] mono-energy}
& \multirow{3}{30mm}{$\vbh = 0$,\\[2pt] mono-energy,\\[2pt] Incomp. $\exb$} \\[2pt]
& & & & &    \\[2pt]
& & & & &    \\[2pt] \hline
$\dot{\mu}$ & 0  & finite & 0 & finite & finite  \\[2pt]
$\dot{v} $  & $\propto -e\Phi'\dot{\psi}$ & $\propto -e\Phi'\dot{\psi}$ & $\propto -e\Phi'\dot{\psi}$  & 0  & 0 \\[2pt]
$\vbh$   & Included  & Included& None   & None & None \\[2pt]
$\nabla\cdot\dot{\bm{z}}$ & 0 & finite & 0 & finite & 0 \\[2pt]
$\bm{E}\times\bm{B}$ drift & Comp. & Comp. & Comp. & Comp. & Incomp. \\[2pt]
 Dimensions & 5  & 4 & 4 & 3 & 3 \\[2pt] \hline
 \end{tabular}
\end{table*}

We derive several kinds of the drift kinetic equation
for the first order distribution function $f_1$
based on various models of the guiding-center particle orbit
stepwisely in the following subsections.
Our starting point is the drift kinetic equation without any assumptions,
which is adopted in the the radially-global neoclassical transport with the FOW effect,
such as GTC-NEO~\cite{Wang2001} and FORTEC-3D~\cite{Satake2008} codes.
The radially local drift kinetic equation is obtained by omitting
the higher order radial drift term, $\dot{s}\partial f_1/\partial s$, from the equation, leading to the ZOW orbit.
Then, further simplifications are usually made to the local drift kinetic equation
in conventional neoclassical transport simulations instead of solving the
ZOW-limit equation directly.
In ZMD limit,
a term involving the tangential magnetic drift to a flux surface is approximated to be zero,
that is, $\vbh\cdot\nabla f_1 = 0$.
In addition,
mono energetic particle $\dot{v}\partial_v f_1 \propto \dot{s}\partial_s f_1 = 0$,
and incompressible $\bm{E}\times\bm{B}$ drift are assumed in DKES-like orbit,
where $v$ is the particle velocity.
Subsidiary changes are also introduced in some essential conservation properties
of the drift kinetic equations along with these assumptions for the local neoclassical transport.
The differences among these global (FOW) and local neoclassical transport models
are summarized in Table~\ref{tb:dke} for the later convenience.

\subsection{Drift kinetic equation with finite orbit width (FOW) effect}
Staring equation is the drift kinetic equation for the guiding-center distribution function, $f_a = f_a(\bm{R}, v, \xi)$:~\cite{Helander2002}
\begin{equation}
  \label{eq:dke}
  \frac{\partial f_a}{\partial t} + \dot{z^j}\delop{f_a}{z^j} = C(f_a),
\end{equation}
where subscript $a$ represents the particle species
and the phase-space variables are denoted by $\bm{z}$ as $z^j = (\bm{R}, v, \xi)$;
$\bm{R}$ is the position vector, and $\xi \equiv v_{\parallel}/v$ is the pitch angle
of the parallel velocity $v_{\parallel} = \bm{v}\cdot\bm{b}$ with the unit vector parallel to the magnetic field,
$\bm{b} = \bm{B}/B$;
$C(f_{a})$ is the linearized collision operator acting on $f_{a}$.
The subscript $a$ is omitted for simplicity hereafter unless it is necessary.
In the following, we use Boozer coordinates~\cite{Boozer1982}
to specify the position vector $\bm{R}$ as $\bm{R} = (\psi, \theta, \zeta)$,
where $\psi$ is the toroidal magnetic flux, $\theta$ and $\zeta$ are the poloidal and toroidal angle variables.
the drift equations of motion are derived from the canonical Hamiltonian given by White:~\cite{White1990}
\begin{subequations}
  \label{eq:gcd fow}
  \begin{align}
    \dot{\theta} &= \frac{1}{\gamma}\left\{G\Phi' + \left(\biota - \frac{mv\xi G'}{eB}\right)v_{\parallel}B +
      \frac{Gmv^2}{2eB}\left(1+\xi^2\right)\delop{B}{\psi}\right\}\\
    \dot{\zeta} &= \frac{1}{\gamma}\left\{-I\Phi'+ \left(1 + \frac{mv\xi I'}{eB} \right)v_{\parallel}B -
      \frac{Imv^2}{2eB}\left(1+\xi^2\right)\delop{B}{\psi}\right\}\\
    \dot{\psi}  &= \frac{mv^2\left(1+\xi^2\right)}{2eB\gamma}
    \left( I\delop{B}{\zeta} - G\delop{B}{\theta}\right)\\
    \dot{v} &= -\frac{v\left(1+\xi^2\right)\Phi'}{2B\gamma}
    \left(I\delop{B}{\zeta} - G\delop{B}{\theta}\right)\\
    \dot{\xi} &= -\frac{1-\xi^2}{2\gamma} \left[
     v\left\{\left(1+\frac{mv\xi I'}{eB}\right)\delop{B}{\zeta}
        + \left(\biota - \frac{mv\xi G'}{eB}\right)\delop{B}{\theta}\right\} 
      + \frac{\xi\Phi'}{B} \left(I\delop{B}{\zeta} - G\delop{B}{\theta}\right) \right],
  \end{align}
\end{subequations}
where $m$ and $e$ are the mass and electric charge of the species;
$\biota(\psi)$ is the rotational transform;
$\Phi = \Phi(\psi)$ is the electrostatic potential;
prime denotes the derivative with respect to $\psi$;
$\gamma = G(1 + \frac{mv\xi}{eB}I') + I(\biota - \frac{mv\xi}{eB}G')$
is used for simplicity
with the poloidal and toroidal current fluxes, $G(\psi)$ and $I(\psi)$.

By introducing a small parameter $\delta \sim {\mathcal O}(\rho/L)$, where $\rho$ represents the Larmor radius
and $L$ denotes the typical scale length,
the drift kinetic equation can be solved order by order.
To this end
two important orderings are assumed;
one is the transport ordering of $\delop{}{t} \sim {\mathcal O}(\delta^2 \omega_{\rm t})$,
where $\omega_{\rm t} \simeq L/v_{\rm th}$ is the transit frequency,
and the other is the drift ordering of $v_{\rm E}/v_{\rm th} \sim \mathcal{O}(\delta)$,
where $\ve$ and $v_{\rm E}$ is the $\exb$ drift and its magnitude,
and $v_{\rm th}$ is the thermal speed of the particle.
For the drift ordering, we use the fact that $\ve$ is tangential to a flux surface
since $\Phi$ is a flux function.
By decomposing the distribution function as $f = f_0 + f_1$,
it can be readily shown that $f_0$ is Maxwellian, that is, $f_0 = f_{\rm M} (\psi, v)$
from the leading order drift kinetic equation.



The drift kinetic equation for $f_1$ becomes as follows:
\begin{eqnarray}
\label{eq:dke f1}
\frac{Df_1}{Dt} &\equiv&
\frac{\partial f_1}{\partial t}
+ \dot{\psi}\frac{\partial f_1}{\partial \psi}
+ \dot{\theta}\frac{\partial f_1}{\partial \theta}
+ \dot{\zeta}\frac{\partial f_1}{\partial \zeta}
+ \dot{v}\frac{\partial f_1}{\partial v}
+ \dot{\xi}\frac{\partial f_1}{\partial \xi} \nonumber \\
&=& -\dot{\psi}\frac{\partial f_0}{\partial \psi} -\dot{v}\frac{\partial f_0}{\partial v} + C(f_1),
\end{eqnarray}
where $f_0 = f_0(\psi, v)$ is used, and $C(f_1) = C(f_1,f_0) + C(f_0,f_1)$
is a linearized collision operator for Coulomb collisions.
It should be noted that eq.\eqref{eq:dke f1} involves terms of different orders.
$\dot{\psi}\partial_\psi f_1$ and $\dot{v}\partial_v f_1$ are of the order of ${\mathcal O}(\delta^2 \omega_{\rm t} f_0)$,
where $\dot{v} \propto \Phi'\dot{\psi}$.
On the other hand, $\dot{\theta} \partial_\theta f_1$, $\dot{\zeta} \partial_\zeta f_1$
and $\dot{\xi} \partial_\xi f_1$ are composed of ${\mathcal O}(\delta \omega_{\rm t} f_0)$ term arising from the parallel motion
and ${\mathcal O}(\delta^2 \omega_{\rm t} f_0)$ terms from the perpendicular drift.
It should be noted that $\partial_t f_1$ is regarded as the order of ${\mathcal O}(\delta^3 \omega_{\rm t} f_0)$
in the quasi-steady state according to the transport ordering.
Solving eq.~\eqref{eq:dke f1} directly with linearized collision operator for $f_1$
leads to the neoclassical transport with the FOW effect which is represented by
the radial drift term $\dot{\psi}$ and velocity term $\dot{v}$.


The drift kinetic equation with the higher order radial drift,
satisfies the conservative properties of phase-space volume and particle number.
This is due to the fact that the guiding-center motion in the five-dimensional phase-space,
$\dot{\bm{z}}$, is Hamiltonian.
For $\dot{\bm{z}}$, Liouville's theorem is satisfied:
\begin{equation}
\label{eq:liouville}
\nabla\cdot\dot{\bm{z}} = \frac{1}{{\cal J}}\sum_{j=1}^5 \left({\cal J}\dot{z}^j\right) = 0,
\end{equation}
where $z^j = (\psi, \theta, \zeta, v, \xi)$,
and the Jacobian of the five-dimensional phase-space coordinates
is given as
\begin{equation}
\label{eq:Jacobian global}
{\cal J} = \frac{2\pi B_{\parallel}^*v^2}{B} \frac{G+\biota I}{B^2},
\end{equation}
with $B^*_{\parallel} \equiv \left(\bm{B} +
\frac{mv\xi}{e}\nabla\times\bm{b}\right)\cdot\bm{b}$.
Here, we assume that the Jacobian does not depend on time explicitly.
Therefore,
the original (global) drift kinetic equation, eq.~\eqref{eq:dke} and \eqref{eq:dke f1},
conserves the five-dimensional phase-space volume.

The particle number in the phase-space is also conserved.
This can be readily seen by rearranging the five-dimensional drift kinetic equation in the conservative form.
The operator ${\cal P}$ is defined as
\begin{eqnarray}
{\cal P} \equiv \delop{}{t} + \frac{1}{\cal J}\delop{}{z^j}{\cal J}\dot{z}^j
= \delop{}{t} + \dot{z}^j \delop{}{z^j}
\end{eqnarray}
where Liouville's theorem, eq.~\eqref{eq:liouville},
and $\partial {\cal J}/\partial t = 0$ are used to show the second equality.
Using ${\cal P}$ and assuming collisionless limit of $C(f) \to 0$,
the drift kinetic equations for $f$ and $f_1$ becomes
\begin{eqnarray}
{\cal P} f &=& 0 \\
{\cal P} f_1 &=& -\dot{\psi}\delop{f_0}{\psi} - \dot{v}\delop{f_0}{v}.
\end{eqnarray}
Integrating the drift kinetic equations over the entire phase-space volume $d\bm{z} = {\cal J}dz^1\cdots dz^5$
and using the definitions above, we have
\begin{eqnarray}
\frac{dN}{dt} &=& 0 \\
\frac{dN_1}{dt} &=& 0.
\end{eqnarray}
In the equations above, the total particle numbers
for $f$ and $f_1$ in the phase-space, are defined by
$N = \int d\bm{z} f$ and $N_1 = \int d\bm{z} f_1$, respectively.
The particle number conservation also holds
for collisional cases since a proper choice of a collision operator
satisfies the conservation laws for the particle number, momentum and energy.~\cite{Satake2008}


\subsection{Local drift kinetic equation in the Zero Orbit Width (ZOW) limit}
The local drift kinetic equation in ZOW limit
is obtained by neglecting the higher order radial drift
term, $\dot{\psi}\partial f_1/\partial \psi$, in eq.~\eqref{eq:dke f1}. We have
\begin{eqnarray}
\label{eq:local dke}
\frac{Df_1}{Dt} &\equiv& \frac{\partial f_1}{\partial t} 
+ \dot{\theta}\frac{\partial f_1}{\partial \theta}
+ \dot{\zeta}\frac{\partial f_1}{\partial \zeta}
+ \dot{v}\frac{\partial f_1}{\partial v}
+ \dot{\xi}\frac{\partial f_1}{\partial \xi} \nonumber \\
&=& -\dot{\psi}\frac{\partial f_0}{\partial \psi}
-\dot{v}\frac{\partial f_0}{\partial v} + C(f_1).
\end{eqnarray}
It should be noticed that $D/Dt$ represents the total derivative
along the particle orbit in $\zlocal = (\theta, \zeta, v, \xi)$.
not in $\bm{z}$ in eq.~\eqref{eq:dke f1}.
What is important in this equation is that
the radial variable $\psi$ only appears in the right hand side of the equation
as a source term, $\dot{\psi}\partial f_0/\partial \psi$.
This means that the dependence on $\psi$ only enters
through $\dot{\psi}\partial f_0/\partial \psi$
as a parameter when equilibrium distribution $f_0$ is given.
The radially local neoclassical transport at a surface can be evaluated independently
by solving eq.~\eqref{eq:local dke} at the surface.
It should be noted that, in ZOW limit,
the higher order radial drift term, $\dot{\psi}\partial_s f_1$, is ignored
while $\dot{v}\partial_v f_1$ term still remains
in order to make the comparison to the further reduced local neoclassical transport model (the ZMD limit) simpler.

The particle orbit lies on a single flux surface during the time evolution.
The drift equations of guiding-center motion for the local drift kinetic equation
are obtained
by omitting the effect of the radial drift in eqs.~\eqref{eq:gcd fow} as
\begin{subequations}
  \label{eq:gcd}
  \begin{align}
  \dot{\theta} &= \frac{1}{G+\biota I}\left\{G\Phi'+\biota v_{\parallel}B + 
  \frac{Gmv^2}{2eB}\left(1+\xi^2\right)\delop{B}{\psi}\right\}\\
  \dot{\zeta} &= \frac{1}{G+\biota I}\left\{-I\Phi'+ v_{\parallel}B -
  \frac{Imv^2}{2eB}\left(1+\xi^2\right)\delop{B}{\psi}\right\}\\
  \dot{v} &= -\frac{v\left(1+\xi^2\right)\Phi'}{2B\left(G+\biota I\right)}
  \left(I\delop{B}{\zeta} - G\delop{B}{\theta}\right)\\
  \dot{\xi} &= -\frac{1-\xi^2}{2\left(G+\biota I\right)} 
  \left\{ v\left(\delop{B}{\zeta} 
   + \biota\delop{B}{\theta}\right) + \frac{\xi\Phi'}{B}
   \left(I\delop{B}{\zeta} - G\delop{B}{\theta}\right) \right\}.
  \end{align}
\end{subequations}
To derive eqs.~\eqref{eq:gcd},
$B^*_{\parallel} \simeq B$ is used,
and the coefficient $\gamma$ is approximated as $ G + \biota I$.
Our main purpose of this paper is to construct a proper numerical method to solve eq.~\eqref{eq:local dke}
along the particle orbit given by eqs.~\eqref{eq:gcd}.


The consequence of the neglect of the radial drift
in the local drift kinetic equation, eq.~\eqref{eq:local dke},
is compressibility of the phase-space volume.
The divergence of the phase-space flow becomes finite
due to the presence of the magnetic drift in poloidal and toroidal directions,
$\vbh$.
In contrast to the case of eq.~\eqref{eq:dke f1},
the conservative form ${\cal P}$ in $\zlocal$ does not agree to the total derivative
along the particle orbit in $\zlocal$-coordinates,
$D/Dt$; ${\cal P} = D/Dt + \nabla\cdot\dzlo$.
In this phase-space, the Jacobian is written as follows:
\begin{equation}
\label{eq:Jacobian local}
{\cal J} = \frac{2\pi B_{\parallel}^*v^2}{B} \frac{G+\biota I}{B^2} \simeq 2\pi v^2 \frac{G+\biota I}{B^2},
\end{equation}
where $B_{\parallel}^*$ is again approximated by $B$
and the $(G+\biota I)/B^2$ part is Jacobian of Boozer coordinates
and it is the same as that of the five-dimensional phase-space.
It should be noted that this approximation of $B_{\parallel}^*$ to $B$ does not influence
on the conservation property of the phase-space volume,
although the Hamiltonian nature of the system is broken.
Using eq.~\eqref{eq:gcd}, the compressibility of the phase-space volume can be obtained as
\begin{eqnarray}
\label{eq:div zdot}
\nabla\cdot\dzlo &=& -\frac{mv^2\left(1+\xi^2\right)}{2eB\left(G+\biota I\right)}
\left\{
\frac{3}{B}\delop{B}{\psi} \left(I\delop{B}{\zeta} - G\delop{B}{\theta}\right) \right. \nonumber \\
&+& \left.
\left(G\frac{\partial^2 B}{\partial \psi \partial \theta} -I\frac{\partial^2 B}{\partial \psi \partial \zeta} \right)
\right\}.
\end{eqnarray}
The right hand side of eq.~\eqref{eq:div zdot}
exclusively arises from $\vbh$ part of $\dot{\theta}$ and $\dot{\zeta}$.
Since $\left(\nabla\cdot\dzlo\right)f_1$ is of the order of ${\mathcal O}(\delta^3)$,
it is a higher order contribution to the local drift kinetic equation,~\eqref{eq:local dke}.
The local drift kinetic equation, \eqref{eq:local dke}, is rewritten as follows:
\begin{equation}
\label{eq:pf1 local}
{\cal P} f_1 = \left(\nabla\cdot\dzlo\right) f_1 + S_0
\end{equation}
where $S_0 = - \dot{v}\delop{f_0}{v} -\dot{\psi}\delop{f_0}{\psi}$
is used to formally represent its behavior as a source term
in the local drift kinetic equation.

Integrating eq.~\eqref{eq:pf1 local}
over the phase space gives rise to the non-vanishing contribution
to $dN_1/dt$:
\begin{equation}
\frac{dN_1}{dt} = \int d\zlocal \left(\nabla\cdot\dzlo\right) f_1,
\end{equation}
where contributions from other source term, $S_0$ 
vanishes since they represent the radial velocity moment of Maxwellian distribution function,
$f_0$.
The number of particle in the phase-space is not conserved in the local drift kinetic equation due to the compressibility
when the finite tangential magnetic drift is considered.
The same situation occurs for the conservation of the magnetic moment $\mu$;
the magnetic drift contributions in $\dot{\theta}$ and $\dot{\zeta}$ again lead to the violation of $\dot{\mu} = 0$:
\begin{equation}
\dot{\mu} = 
 \frac{mv_{\perp}\dot{v}_{\perp}}{B}
-\frac{mv_{\perp}^2}{2B^2}\left(\dot{\theta}\delop{B}{\theta} + \dot{\zeta}\delop{B}{\zeta}\right) 
= \frac{\mu}{B}\delop{B}{\psi}\dot{\psi}.
\end{equation}

Although the phase-space volume, and thus the particle number $N_{1}$ are not conserved
when considering the ZOW-limit particle orbit,
this does not cause any matter practically in evaluating steady-state neoclassical transport observables
such as the particle and energy fluxes in many cases
by the $\delta f$ Monte Carlo method prescribed in Sec.~\ref{sec:method}.
In fact, the neoclassical transport observables presented in Secs.~\ref{sec:axisym} and \ref{sec:assym}
reach steady-state values in our particle simulations while $N_1$ remains negligible compared to $N$.
The ZOW model is inappropriate only when extremely large radial excursion of the guiding centers,
such as the potato orbit near the axis,~\cite{Lin1997a,Satake2002}
mainly determines the neoclassical transport.
In fact, all the local models of the neoclassical transport are insufficient in such cases,
and the global, or the neoclassical transport with the FOW effect is essentially required.
It should be also noted that, although $\nabla\cdot\dzlo$ acts as a source term in the local drift kinetic equation,
the term is quite different from the source term of Landreman {\it et al}.~\cite{Landreman2014}
It is pointed out in the reference that,
for the cases of the drift kinetic equation with the {\it full} and {\it partial} trajectories,
surface averaged conservation laws of the particle number and energy
result in a singular perturbation problem when $E_r$ approaches to $0$,
where the full trajectory corresponds to the ZMD orbit in this paper.
Their source term is introduced to remove the singular perturbation.
On the other hand, $\nabla\cdot\dzlo$ term is introduced here to
solve the local drift kinetic equation with the non-Hamiltonian property by particle simulations.

The radial locality and the requirement of the divergence-free
phase-space flow do not hold simultaneously.
It should be noticed that this cannot be avoided even if one chooses
other variables for the velocity space.
For example, when we chooses $(v_{\parallel}, \mu)$ as independent variables instead of $(v,\xi)$ and keep $\dot{\mu}=0$,
then the conservation of the total energy along the particle orbit does not hold.
Nevertheless, as presented later in Sec.~\ref{sec:method},
one can numerically solve the local drift kinetic equation
by treating the compressibility as another source term in a system
as well as $S_0$.

\subsection{Local drift kinetic equation in the Zero Magnetic Drift (ZMD) limit}
In some neoclassical transport simulations, additional assumption is made for the local drift kinetic equation
in ZOW limit presented in the previous subsection;
$\vbh\cdot\nabla f_1 = 0$ is assumed to neglect the tangential magnetic drift.
Among them are EUTERPE code~\cite{Garcia-Regana2013a} and the works of Landreman,~\cite{Landreman2013,Landreman2014} for example.
Hence we refer to this type of the particle orbit as a ZMD orbit
to distinguish it from the ZOW orbit and DKES-like orbit in this paper.

By omitting the magnetic drift from eq.~\eqref{eq:gcd},
the equations of the guiding-center drift are given as
\begin{subequations}
  \label{eq:gcd no vb}
  \begin{align}
  \dot{\theta} &= \frac{1}{G+\biota I}\left(G\Phi'+\biota v_{\parallel}B\right)\\
  \dot{\zeta} &= \frac{1}{G+\biota I}\left(-I\Phi'+ v_{\parallel}B\right)\\
  \dot{v} &= -\frac{v\left(1+\xi^2\right)\Phi'}{2B\left(G+\biota I\right)}
  \left(I\delop{B}{\zeta} - G\delop{B}{\theta}\right)\\
  \dot{\xi} &= -\frac{1-\xi^2}{2\left(G+\biota I\right)} 
   \left\{ v\left(\delop{B}{\zeta} 
   + \biota\delop{B}{\theta}\right) + \frac{\xi\Phi'}{B}
   \left(I\delop{B}{\zeta} - G\delop{B}{\theta}\right) \right\}
  \end{align}
\end{subequations}
The local drift kinetic equation is formally the same as eq.~\eqref{eq:local dke}.
Solving eq.~\eqref{eq:local dke} along with this modified guiding-center orbit,
eqs.~\eqref{eq:gcd no vb}, leads to the local neoclassical transport without $\vbh$.

$\nabla\cdot\dzlo = 0$ is again recovered
due to the absence of the magnetic drift terms in $\dot{\theta}$ and $\dot{\zeta}$
and the presence of $\dot{v}$ term in this limit.
(Neglecting $\dot{v}$ term again gives rise to the finite $\nabla\cdot\dzlo$, see eq.~\eqref{eq:comp 3d}.)
This leads to the conservation of the particle number $N_1$
since conservative form ${\cal P}$ agrees with the total derivative along the orbit, $D/Dt$.
The conservation of $\mu$ is also followed from the absence of $\vbh$.
It should be noted that,
although the magnetic drift terms in $\dot{\theta}$ and $\dot{\zeta}$ are
of the order of $\delta^2$ as well as the $\exb$ drift,
the latter is only taken into account in this limit.
This causes a large peak of neoclassical radial fluxes around $\Er = 0$.


\subsection{Mono-energetic assumption and incompressible $\exb$ drift (DKES-like limit)}
In this subsection, the local drift kinetic equation in the ZMD limit is further
reduced by assuming the so-called mono-energetic particles ($\dot{v} = 0$),
in which the kinetic energy does not experience any change.
As shown later in this subsection,
the assumption of the mono-energetic particle again violates the phase-space volume conservation.
To recover the conservation property, the $\exb$ drift is also assumed to be incompressible,
resulting in the DKES-like limit particle orbit.
It should be noted that
we always use the DKES-like limit ($\dot{v} = 0$ and incompressible $\exb$ drift)
when considering the mono-energetic particle assumption in this paper.

When the mono-energetic assumption, $\dot{v}\partial f_1/\partial v = 0$, is made
in addition to the assumption of $\vbh\cdot\nabla f_1 = 0$,
the local drift kinetic equation reduces to three-dimensional problem,
Since $\dot{v}$ term is higher order as described above,
this assumption is consistent to neglecting the higher order radial drift.
Under the mono-energetic assumption,
the local drift kinetic equation~\eqref{eq:local dke} becomes
\begin{equation}
\label{eq:local dke mono-e}
\frac{\partial f_1}{\partial t} + \dot{\theta}\frac{\partial f_1}{\partial \theta}
+ \dot{\zeta}\frac{\partial f_1}{\partial \zeta} 
+ \dot{\xi}\frac{\partial f_1}{\partial \xi}
= S_0 + C(f_1).
\end{equation}
In the equation, the particle velocity $v$ (kinetic energy $mv^2/2$)
only enters parametrically through $\partial f_0/\partial v$ in the right hand side.
Also, the test-particle collision operator $C_{\rm T}(f_1)$
included in the linearized collision operator $C(f_1)$
should be also modified under the mono-energetic assumption.
The test-particle collision operator is reduced
to the pitch-angle scattering operator (Lorentz operator).
The particles do not experience the energy scattering.
The four dimensional phase space $\zlocal$ reduces to
three-dimensional one, ${\bm z}^{(3)} = (\theta, \zeta, \xi)$.
$v$ can be treated just as a parameter to solve the equation
as well as the radial variable $s$.
The mono-energetic guiding-center drift equations of motion is
the same as eqs.~\eqref{eq:gcd no vb}
except for $\dot{v} = 0$ in this case.

The mono-energetic assumption again violates
phase-space volume and particle number conservations.
This arises due to the presence of the $\exb$ drift
in $\dot{\theta}$, $\dot{\zeta}$ and $\dot{\xi}$.
The divergence of $\dot{\bm{z}}$ becomes
\begin{equation}
\label{eq:comp 3d}
\nabla\cdot\dot{\bm{z}}^{(3)} = \frac{1}{{\cal J}}\frac{3\left(1+\xi^2\right)}{2B^3}
\frac{d\Phi}{d\psi}\left(G\delop{B}{\theta} - I\delop{B}{\zeta}\right),
\end{equation}
where ${\cal J}$ is Jacobian given in eq.~\eqref{eq:Jacobian local}.
$\exb$ drift also results in $\dot{\mu} \propto d\Phi/d\psi$;
the magnetic moment is not conserved along the mono-energetic guiding-center orbit.


$\exb$ drift is often regarded as an incompressible drift in conventional local neoclassical transport codes,
such as DKES.~\cite{Hirshman1986,VanRij1989}
The $\exb$ drift, is approximated as follows:
\begin{equation}
\frac{\exb}{B^2} \to \frac{\exb}{\langle B^2\rangle},
\end{equation}
or equivalently,
$\Phi'$ in the guiding-center drift equations of motion is replaced by $\Phi'B^2/\langle B^2 \rangle$.
With this replacement,
the guiding-center drift equations of motion then becomes
\begin{subequations}
  \label{eq:gcd exb1}
  \begin{align}
  \dot{\theta} &= \frac{1}{G+\biota I}\left(G\frac{B^2}{\langle B^2\rangle}\Phi'
    +\biota v_{\parallel}B\right)\\
  \dot{\zeta} &= \frac{1}{G+\biota I}\left(-I\frac{B^2}{\langle B^2\rangle}\Phi'
    + v_{\parallel}B\right)\\
  \dot{\xi} &= -\frac{1-\xi^2}{2\left(G+\biota I\right)} 
   \left\{ v\left(\delop{B}{\zeta} 
   + \biota\delop{B}{\theta}\right) + \frac{\xi\Phi'}{B}
   \left(I\delop{B}{\zeta} - G\delop{B}{\theta}\right) \right\}
  \end{align}
\end{subequations}
where $\vbh = 0$ and $\dot{v} = 0$ are also assumed.
While the local drift kinetic equation along this guiding-center orbit
still violates the phase-space volume conservation,
it is satisfied if effect of $\Phi'$ is simultaneously removed from $\dot{\xi}$.
Indeed, this is what DKES and many conventional neoclassical transport code assume
in their approach; no magnetic drift, mono-energetic particle,
incompressible $\exb$ drift, and no $\Phi'$ effect on $\dot{\xi}$.
We call this particle orbit DKES-like orbit, for simplicity.

Adopting these all assumptions,
the drift equations of motion for DKES-like orbit are
\begin{subequations}
  \label{eq:gcd exb2}
  \begin{align}
  \dot{\theta} &= \frac{1}{G+\biota I}\left(G\frac{B^2}{\langle B^2\rangle}\Phi'
    +\biota v_{\parallel}B\right)\\
  \dot{\zeta} &= \frac{1}{G+\biota I}\left(-I\frac{B^2}{\langle B^2\rangle}\Phi'
    + v_{\parallel}B\right)\\
  \dot{\xi} &= -\frac{v\left(1-\xi^2\right)}{2\left(G+\biota I\right)}
  \left(\delop{B}{\zeta} + \biota\delop{B}{\theta}\right)
  \end{align}
\end{subequations}
As a consequence, the phase-space volume conservation is again satisfied
with $\nabla\cdot\dot{\bm{z}}^{(3)} = 0$.
Hereafter, in this paper, the guiding-center particle orbit with incompressible $\exb$ drift
represents those described by eq.~\eqref{eq:gcd exb2}, not eq.~\eqref{eq:gcd exb1}.
On the other hand, however,
the violation of $\dot{\mu} = 0$ is not recovered
even if incompressible $\exb$ drift is assumed;
$\mu$ is not conserved along the guiding-center trajectory due to the radial electric field.

These additional assumptions are simultaneously adopted
in many conventional neoclassical transport codes.
This makes it difficult to properly compare the difference
among various models of the local drift kinetic equation and/or
the drift kinetic equation with FOW effect.
In order to address the effect of each drift on the local neoclassical transport,
it is necessary to construct a numerical method to
solve the wide varieties of the local drift kinetic equations with
several drifts included/neglected independently.

\section{2-weight $\delta f$ Monte Carlo method for local neoclassical transport}
\label{sec:method}
Two-weight $\delta f$ Monte Carlo method is widely used
to solve the drift kinetic equation and its formulation for collisional transport
with incompressible flow of $\nabla\cdot\bm{z} = 0$
was given in detail by
Brunner {\it et al.}~\cite{Brunner1999}
and Wang {\it et al.}~\cite{Wang1999a} respectively.
Since our interest is the local drift kinetic equation in ZOW limit,
where $\nabla\cdot\dzlo \ne 0$,
the formulation needs to be modified to appropriately treat such case.

Hu and Krommes pointed out in their work,~\cite{Hu1994}
the two-weight $\delta f$ Monte Carlo method is applicable to a non-Hamiltonian
system in which the compressibility of the phase-space volume is included as a source term
in the weight evolutions.
According to the work,
we apply the method to the local drift kinetic equation in ZOW limit.
Below, we briefly review the standard formulation
of the two-weight $\delta f$ Monte Carlo method for Hamiltonian (incompressible flow)
system.
The formulation is given for five-dimensional phase-space coordinates for generality.
Then, to discuss the ZOW limit in four-dimensional case,
the effect of $\nabla\cdot\dzlo$ term is included as a source term.
Cases of ZMD and DKES-like limit are then presented.


In the two-weight method,
two weights, $w$ and $p$, are assigned to each simulation marker.
Then, the discretized distribution function of simulation markers,
$F = F(\bm{z}, w, p; t)$, is introduced.
It is noted that the distribution function $F$ is defined
not in an ordinary phase-space $\bm{z}$, but in an extended phase-space $(\bm{z}, w, p)$.
Using $F$, the distribution functions, $f_0$ and $f_1$, are evaluated
weighted sum of $F$ as follows:
\begin{subequations}
  \label{eq:marker_dist}
  \begin{align}
    \label{eq:marker_dist1}
    F &= \sum_i \delta(\bm{z}-\bm{z}_i)\delta(w-w_i)\delta(p-p_i) {\cal J}^{-1}(\bm{z})\\
    \label{eq:marker_dist2}
    g &= \int F dwdp = \sum_i \delta(\bm{z}-\bm{z}_i) {\cal J}^{-1}(\bm{z}) \\
    \label{eq:marker_dist3}
    f_0 &= \int pF dwdp = \sum_i p_i \delta(\bm{z}-\bm{z}_i) {\cal J}^{-1}(\bm{z})\\
    \label{eq:marker_dist4}
    f_1 &= \int wF dwdp = \sum_i w_i \delta(\bm{z}-\bm{z}_i) {\cal J}^{-1}(\bm{z}),
  \end{align}
\end{subequations}
where $g = g(\bm{z})$ denotes the marker distribution function
in the ordinary phase space $\bm{z}$,
and subscript $i$ denotes the marker indices.
The expressions for the weights $w_i$ and $p_i$ are obtained 
by integrating eqs.~\eqref{eq:marker_dist} for $f_0$ and $f_1$
using $F$ and $g$:
\begin{subequations}
  \label{eq:wp_def}
  \begin{align}
    w_i =& \frac{f_1(\bm{z}_i)}{g(\bm{z}_i)} \\
    p_i =& \frac{f_0(\bm{z}_i)}{g(\bm{z}_i)}.
  \end{align}
\end{subequations}

The total derivative along the particle orbit including the test-particle collision
$D^{(c)}/Dt$ is defined by rewriting eq.~\eqref{eq:dke f1} as
\begin{eqnarray}
\label{eq:dke f1 2}
\frac{D^{(c)}f_1}{Dt} &\equiv& \frac{Df_1}{Dt} - C_{\rm TP}(f_1) \nonumber \\
&=& S_0 +  C_{\rm FP}(f_{\rm M}),
\end{eqnarray}
where the linearized collision operator $C(f_1)$ is decomposed into 
the test-particle part, $C_{\rm TP}(f_1)$ and field-particle one, $C_{\rm FP}(f_{\rm M})$.
Since the simulation markers are discretized,
the total derivative along the particle $D^{(c)}/Dt$ should be replaced by
what is appropriate for the discretized markers.
In the two-weight $\delta f$ Monte Carlo method, this is enabled
by approximating the test-particle collision in $D^{(c)}/Dt$
by Monte Carlo collision operator for the discretized markers;~\cite{Xu1991,Lin1995}
$D^{({\rm M})}/Dt \simeq D^{(c)}/Dt$.
It is worth noting that this approximation of the collision operator is the origin
of the weight spreading.~\cite{Brunner1999}

The marker distribution function $g(\bm{z})$ is conserved along $D^{(c)}/Dt$:
\begin{equation}
\label{eq:dke g}
\frac{D^{(c)}g}{Dt} = {\cal P}g -g\left(\nabla\cdot\dot{\bm{z}}\right) -C_{\rm TP}(g) = 0,
\end{equation}
since $\nabla\cdot\dot{\bm{z}} = 0$.
Thus, we obtain following equation from eq.~\eqref{eq:dke f1 2},
\begin{equation}
\label{eq:dke g 2}
\frac{D^{(c)}f_1}{Dt} = w\frac{D^{(c)}g}{Dt} + g\frac{D^{(c)}w}{Dt} = g\frac{D^{(c)}w}{Dt}.
\end{equation}
Using $D^{(c)}/Dt \simeq D^{({\rm M})}/Dt$ for the right hand side of the second equality,
the time evolution of $w$ along the marker orbit with the Monte Carlo collision
is obtained as
\begin{equation}
  \label{eq:evo w}
  \dot{w}_i \equiv \frac{D^{({\rm M})}w}{Dt} = -\frac{p_i}{f_{\rm M}} \left(\dot{\psi}\frac{\partial}{\partial \psi} + \dot{v}\frac{\partial}{\partial v} - C_{\rm FP} \right) f_{\rm M},
\end{equation}
where eq.~\eqref{eq:dke f1 2} is used for the right hand side.
Similarly, time evolution of $p$ can be read as
\begin{equation}
  \label{eq:evo p}
  \dot{p}_i = \frac{p_i}{f_{\rm M}}
\left(\dot{\psi}\frac{\partial}{\partial \psi} + \dot{v}\frac{\partial}{\partial v} \right) f_{\rm M}.
\end{equation}



When we consider the local drift kinetic equation in ZOW limit in $\zlocal$ phase space,
the compressibility of the phase-space volume remains in eq.~\eqref{eq:dke g}.
Hereafter in this section,
we use formally the same notations for the distribution functions
$f_1$, $g$, {\it etc.}, although they are defined in $\zlocal$ phase space not
in $\bm{z}$.
According to this change, the total derivatives $D^{(c)}/Dt$ and $D^{(M)}/Dt$ also
become those defined in $\zlocal$.

With finite $\nabla\cdot\dzlo$,
the marker distribution along $D^{(c)}/Dt$ becomes
\begin{equation}
\label{eq:local dke g}
\frac{D^{(c)}g}{Dt} = -g\left(\nabla\cdot\dzlo\right).
\end{equation}
For eq.~\eqref{eq:local dke},
a similar discussion as in eqs.~\eqref{eq:dke f1 2} - \eqref{eq:dke g 2} leads
to the time evolution of $w$ as follows:
\begin{eqnarray}
  \label{eq:evo w local}
  \dot{w}_i &=& -\frac{p_i}{f_{\rm M}} \left(\dot{\psi}\frac{\partial}{\partial \psi} + \dot{v}\frac{\partial}{\partial v} - C_{\rm FP} \right) f_{\rm M}
+ w\left( \nabla\cdot\dzlo \right)
\end{eqnarray}
To obtain the time evolution of $p$, it should be noticed that
the total derivative $D^{(c)}/Dt \simeq D^{(M)}/Dt$
is described in $\zlocal = (\theta, \zeta, v, \xi)$.
For $p$ we obtain
\begin{eqnarray}
  \label{eq:evo p local}
  \dot{p}_i &=& \frac{1}{g}\frac{D^{(c)}f_{\rm M}}{Dt} + p\left( \nabla\cdot\dzlo \right)
\nonumber \\
&=& \frac{p_i}{f_{\rm M}} \dot{v}\frac{\partial f_{\rm M}}{\partial v}
+ p\left( \nabla\cdot\dzlo \right).
\end{eqnarray}
Thus, only $v$ derivative and compressibility appear in the right hand side.
The solution of the local drift kinetic equation \eqref{eq:local dke} is obtained
by following the time evolution along the orbit defined
by eqs.~\eqref{eq:gcd} with the Monte Carlo test-particle collision.


The source term of the phase-space incompressibility
is of the order of $f_1\left(\nabla\cdot\dzlo\right) \sim 
w\left(\nabla\cdot\dzlo\right) \sim  {\mathcal O}(\delta^2 \omega_{\rm t}f_0)$.
This means that such a non-conservative property introduced
to the local drift kinetic equation induces higher order effect on the neoclassical transport
as the $\exb$ drift, mono-energetic particle assumption, {\it etc.}
Thus, the use of the finite $\vbh$ can be justified
in solving the local drift kinetic equation, eq.~\eqref{eq:local dke},
which is of the order of ${\mathcal O}(\delta \omega_{\rm t} f_0)$.
The violation of the conservation of the phase-space volume occurs in the local neoclassical transport calculations,
if we would like to treat the finite $\vbh$ in the local drift kinetic equation (ZOW limit).

To avoid the phase-space volume compressibility in the $\zlocal$,
$\vbh = 0$ must be assumed.
The phase-space volume is conserved along the orbit, that is, $\nabla\cdot\dzlo = 0$.
The second terms in
the right hand side of eq.~\eqref{eq:evo w local} and \eqref{eq:evo p local}
reduce to zero in the ZMD limit.

Finally, in DKES-like limit,
same arguments are made in 
three-dimensional phase-space coordinates, $\bm{z}^{(3)} = (\theta, \zeta, \xi)$,
due to the mono-energy assumption.
Noting that $\nabla\cdot\dot{\bm{z}}^{(3)} = 0$ in this phase space,
time evolutions of the weights are described as
\begin{eqnarray}
  \label{eq:evo w and p DKES}
  \dot{w}_i &=& -\frac{p_i}{f_{\rm M}} \left(\dot{\psi}\frac{\partial}{\partial \psi} + \dot{v}\frac{\partial}{\partial v} - C_{\rm FP} \right) f_{\rm M} \\
  \dot{p}_i &=& 0,
\end{eqnarray}
where the time evolution of $p$ can derived by
using the fact that the total derivative $D^{(c)}/Dt$ is described
in $\bm{z}^{(3)}$.
The second weight $p_i$ of each simulation marker is conserved during a simulation.
The two-weight $\delta f$ method in DKES-like limit can be interpreted as
the one-weight method due to the mono-energetic particle.

\section{Axisymmetric case}
\label{sec:axisym}
As a code verification,
several neoclassical transport values are compared to theoretical estimations
for an axisymmetric configuration.
To see the difference among the particle orbit discussed in Sec.~\ref{sec:dke},
we use three kinds of the particle orbit and collision;
one is DKES-like orbit with the pitch angle scattering and without the field-particle
collision operator (denoted as DKES-like, PAS);
the second one also has the DKES-like particle orbit
with full test-particle collision operator
and field-particle collision operator (DKES-like, FC);
the particle orbit of the third one is ZOW orbit with full test- and field-particle
collision operators (ZOW, FC).
For the reference against the radially-global neoclassical code,
numerical simulations are also performed by using the global code, FORTEC-3D (F3D).
It should be noted that the same full collision operator including the field particle operator
as that described above is used in the global code FORTEC-3D.

\begin{figure}[t]
  \centering
   \includegraphics[width=0.4\textwidth]{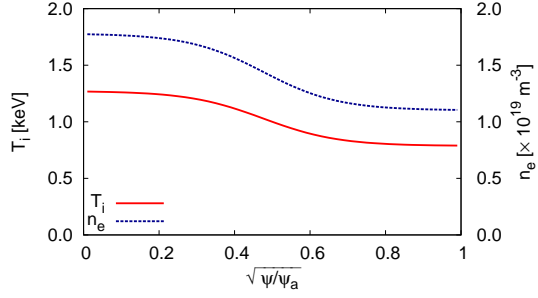}
  \caption{The radial profiles of
    and ion temperature ($T_{\rm i}$) and the plasma density ($n_{\rm e}$).}
      \label{fig:param_tok1}
\end{figure}

\begin{figure}[t]
  \centering
   \includegraphics[width=0.4\textwidth]{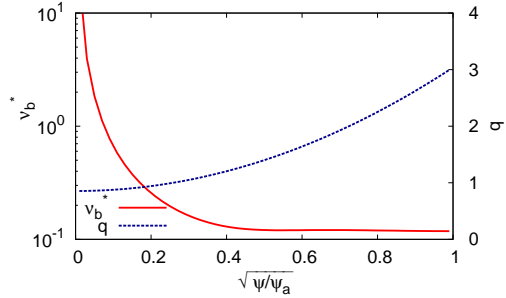}
  \caption{The radial profile of the banana-normalized collisionality $\nu_{\rm b}^*$
    and the safety factor $q$.}
s  \label{fig:param_tok2}
\end{figure}

We use an axisymmetric tokamak geometry with a circular cross-section, of which equilibrium is
constructed by VMEC code~\cite{Hirshman1991} with parameters below;
$R_0$ and $a$ are
the magnetic axis and minor radius are given as
$R_0 = 2.35 ~{\rm m}$ and $a = 0.47 ~{\rm m}$, respectively;
the aspect ratio at the plasma edge $\epsilon^{-1} \equiv R_0/a = 5.0$;
the magnitude of the magnetic field at the magnetic axis is $B_0 =  1.91 ~{\rm T}$.
The safety factor $q = 0.854 + 2.184 \rho^2$ is used.
Hereafter we use
the normalized toroidal magnetic flux, $\sqrt{\psi/\psi_{\rm a}}$, as a flux-surface label,
where $\psi_{\rm a}$ is the toroidal magnetic flux at the plasma edge.
The plasma density $n$ and ion temperature $T_{\rm i}$ are shown
in Fig.~\ref{fig:param_tok1}.
The normalized collisionality, $\nu_{\rm b}^*$, is shown
in Fig.~\ref{fig:param_tok2},
where $\nu_{\rm b}^*$ is the normalized collisionality defined $\nu_{\rm b}^* \equiv
qR/(v_{\rm th, i}\epsilon^{3/2}\tau_{\rm ii})$;
$v_{\rm th, i} = \sqrt{2T_{\rm i}/m_{\rm i}}$ denotes the thermal velocity of the ion;
$\tau_{\rm ii}$ represents the ion-ion collision time~\cite{Hinton1976}
defined from the ion collision time of Braginskii~\cite{Braginskii1965} $\tau_{\rm i}$ as $\tau_{\rm i} = \sqrt{2}\tau_{\rm ii}$
In the Fig.~\ref{fig:param_tok2} (b), the safety factor $q$ is also represented.
The radial electric field is set to be constant during a simulation
and is given as a numerical parameter
according to the force balance relation:~\cite{Hinton1976}
\begin{eqnarray}
\label{eq:force balance}
\langle V_{{\rm i}, \parallel} B\rangle 
&=& \frac{qGT_{\rm i}}{e}\left(\frac{d\psi}{dr}\right)^{-1} 
 \left\{
    \left(\beta_{\rm i}^{\rm NC}-1\right)\frac{d\ln T_{\rm i}}{dr} - \frac{d\ln n_{i}}{dr}
    -\frac{e}{T_{\rm i}}\frac{d\Phi}{dr}\right\},
\end{eqnarray}
where the ion parallel flow $V_{\parallel} = 0$ is assumed to be zero,
and the coefficient $\beta_{\rm i}^{\rm NC}$ is also given
in eqs.(6.134) and (6.135) in the reference.~\cite{Hinton1976}
It should be noted that $\beta_{\rm i}^{\rm NC}$ depends on the collisionality;
since the collisionality is artificially varied to see the collisionality dependence in numerical results presented below,
the radial electric field is also varied depending on the collisionality.
The $\Er$ determined as such enables us to obtain the steady-state neoclassical transport
with less computational time.
This does not affect numerical results
as the neoclassical transport in an axisymmetric tokamak is independent of $\Er$.

\begin{figure}[t]
  \centering
  \includegraphics[width=0.4\textwidth]{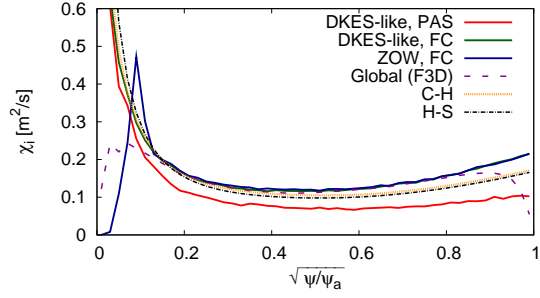}
  \caption{The radial profile of the ion thermal diffusivity obtained by simulations
    with several kinds of the particle orbit and collision.
    DKES-like denotes the particle orbit in the DKES-like limit, and ZOW is that in the Zero Orbit Width limit,
    and Global represents the radially global orbit obtained by FORTEC-3D.
    PAS represents that it uses only the pitch angle collision and does not include  field particle operator;
    FC represents that the simulation uses the full collision operator
    composed of the test particle collision operator with the pitch angle and energy scattering
    and the field particle operator.
    Theoretical estimations from Chang-Hinton (C-H) formula
    and the moment method of Hirshman-Sigmar (H-S) are
    also shown by dotted and chain lines, respectively.}
  \label{fig:chii_cm1}
\end{figure}

\begin{figure}[t]
  \centering
  \includegraphics[width=0.4\textwidth]{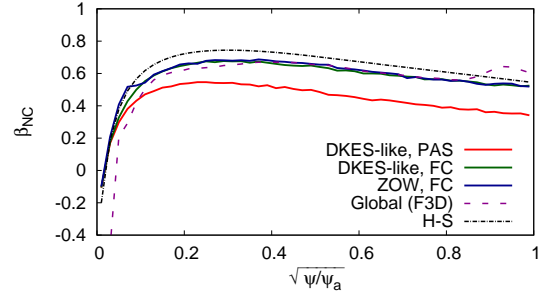}
  \caption{The radial profile of $\beta_{\rm i}^{\rm NC}$ obtained by simulations
    with several kinds of the particle orbit.
    Abbreviation for the kinds of the orbit and collision is the same in Fig.~\ref{fig:chii_cm1}.
    Theoretical estimation of Hirshman-Sigmar (H-S) formula is also shown by a chain line.}
  \label{fig:beta_cm1}
\end{figure}

The radial profile of the neoclassical ion thermal diffusivity $\chi_{\rm i}$ is compared
to theoretical estimates from Chang-Hinton formula~\cite{Chang1982}
and the moment method of Hirshman-Sigmar~\cite{Hirshman1981,Sugama2008} in Fig.~\ref{fig:chii_cm1}.
It should be noticed that the theoretical estimations
for ion thermal diffusivity are obtained
by imposing a certain limitation on the aspect ratio on the local drift kinetic equation
along with the mono-energetic particles and zero tangential magnetic drift.
It is demonstrated that the numerical results of ZOW and DKES-like orbit
with full collision operator (FC) well reproduce the theoretical results over the wide region of the plasma,
while the DKES-like orbit only with pitch angle scattering (without field-particle part)
tends to underestimate the thermal conductivity.
The result shows that the momentum-conservation of the collision operator more influences
neoclassical transport simulations than the difference in the particle orbits.
On the other hand, $\vbh$ hardly influences on the neoclassical thermal transport.
This is because trapped particles with $\vbh$ in the axisymmetric tokamak just precess
in the symmetry direction, and this causes no additional neoclassical transport.

The results of the DKES-like orbit with both collision cases (FC and PAS)
increase towards the magnetic axis of $\sqrt{\psi/\psi_{\rm a}} < 0.15$ as Chang-Hinton and Hirshman-Sigmar theories predict.
On the other hand, the result of the ZOW orbit show decreases towards the axis after
showing an unphysical increase around there.
This suggests that the incompressibility of the phase-space volume in the ZOW limit,
which is caused by the radial deviation of the particle, becomes significant.
In fact, $\chi_{\rm i}$ of the FOW case (denoted as global in the figure)
also shows a discrepancy from those of the DKES-like orbit cases and theoretical values
near the axis, where $\chi_{\rm i}$ smoothly decreases towards zero.
The decreasing tendency towards the axis can be attributed to the existence of the potato orbit
which has the large radial deviation near the axis of a tokamak.~\cite{Lin1997a,Satake2002}
The width of the potato orbit becomes $\simeq 0.13$ for the parameters used here,
showing that the FOW effect due to the orbit becomes significant $\sqrt{\psi/\psi_{\rm a}} < 0.13$.
In other words,
the compressibility of the phase-space volume introduced by the radial drift results in the unphysical transport there.
It should be noted that
$\chi_{\rm i}$ of the global case shows a decrease in the edge region
since the particle loss at the last closed flux surface is only included in the radially global simulation.
Also, both theoretical predictions underestimate $\chi_{\rm i}$ towards the edge region
due to the effect of the finite aspect ratio,
which is only partly included in the theoretical calculation.

The radial profile of $\beta_{\rm i}^{\rm NC}$ is shown in Fig.~\ref{fig:beta_cm1}.
In the figure, $\beta_{\rm i}^{\rm NC}$, is compared to theoretical values of Hirshman-Sigmar.~\cite{Hirshman1981}
It should be noticed that
the momentum conservation does not hold
for the result of the DKES-like PAS case due to the absence of the field particle operator is not included
in the simulations.
The cases with full collision operator with both local orbits (DKES-like and ZOW)
again show a better agreement with theoretical values.

\begin{figure}[t]
  \centering
  \includegraphics[width=0.45\textwidth]{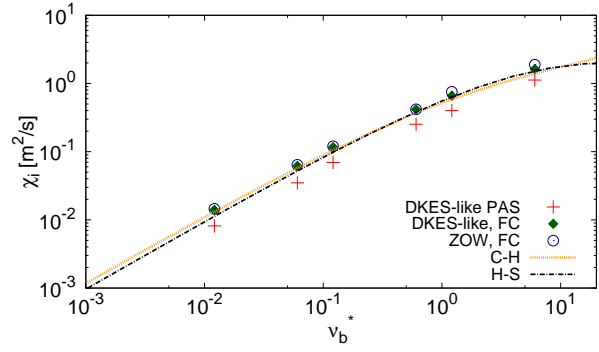}
  \caption{Collisionality dependence of $\chi_{\rm i}$ at $\rho \simeq 0.49$.
    The same abbreviation as in Fig.~\ref{fig:chii_cm1} is used for the kinds of the orbit and collision.
    Chang-Hinton (C-H) and Hirshman-Sigmar (H-S) estimations are represented by dotted and chain lines, respectively.}
  \label{fig:nudep1}
\end{figure}

\begin{figure}[t]
  \centering
  \includegraphics[width=0.45\textwidth]{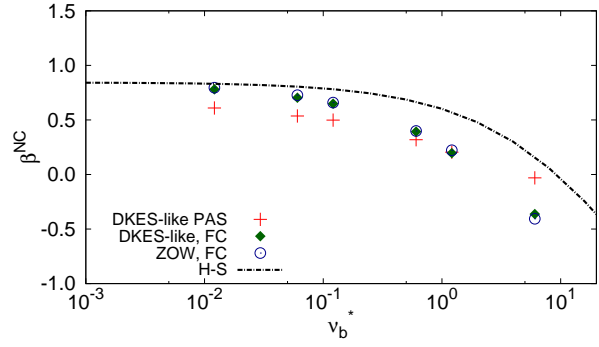}
  \caption{Collisionality dependence of $\beta_{\rm i}^{\rm NC}$ at $\rho \simeq 0.49$.
    The same abbreviation as in Fig.~\ref{fig:chii_cm1} is used for the kinds of the orbit and collision.
    A chain line represents the Hirshman-Sigmar (H-S) formula.}
  \label{fig:nudep2}
\end{figure}

Then,
the collisionality dependence of $\chi_{\rm i}$ and $\beta_{\rm i}^{\rm NC}$
at $\sqrt{\psi/\psi_{\rm a}} \simeq 0.49$ is investigated.
For this purpose,
the collisionality in Fig.~\ref{fig:param_tok2} is numerically magnified by
$0.01$, $0.5$, $5$ and $10$.
It should be noted  that $\Er$ given for each collisionality case
is varied due to the difference of $\beta_{\rm i}^{\rm NC}$ at the initial state
as described before.
The results of $\chi_{\rm i}$ and $\beta_{\rm i}^{\rm NC}$ are shown in
Figs.~\ref{fig:nudep1} and \ref{fig:nudep2}, respectively.
It is shown that the numerical results of the ZOW and DKES-like orbit cases
with the full collision operator (FC)
show better agreement with both theoretical values over the wide range of the collisionality.
$\beta_{\rm i}^{\rm NC}$ of the ZOW and DKES-like orbits reproduce
Hirshman-Sigmar estimates,
especially in the low collisionality regime of $\nu^*_{\rm b} < 0.1$.

\section{non-axisymmetric case}
\label{sec:assym}
To see the effect of the tangential magnetic field drift,
the neoclassical transport in an non-axisymmetric magnetic field configuration
is investigated.
For this purpose, we take an LHD configuration as an example.
Due to the asymmetry in the magnetic field,
the intrinsic ambipolar condition is broken;
the neoclassical transport depends on the radial electric field.
The dependence of the neoclassical transport on $\Er$ shows a resonant peak
at a finite $\Er$ when the tangential magnetic drift exists,
while the peak appears at $\Er = 0$ without the tangential drift.~\cite{Matsuoka2011a}
However, this was demonstrated by comparing local codes (GSRAKE~\cite{CDBeidlerandWDD'haeseleer1995,Beidler2001} and DCOM/NNW~\cite{Wakasa2007}) in DKES limit
and a global code (FORTEC-3D), in which the FOW effect and the tangential drift
was both included.
The role of the tangential magnetic drift in the local drift kinetic equation
is numerically studied below
by using the local code developed here with the ZOW, ZMD and DKES-like orbits.
The validity of the numerical results are also checked
by comparing the results to conventional local neoclassical codes and FORTEC-3D.

\begin{figure}[t]
  \centering
   \includegraphics[width=0.4\textwidth]{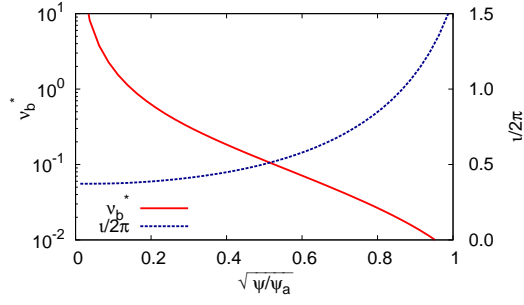}
  \caption{The radial profile of the banana-normalized collisionality $\nu_{\rm b}^*$
    and the rotational transform $\biota$.}
  \label{fig:param_lhdTe1n05}
\end{figure}

The so-called inward-shifted magnetic field configuration of LHD is chosen as
magnetic axis $R_{\rm ax} = 3.6 ~{\rm m}$,
and the magnetic field strength at the axis is $B_{\rm ax} = 3.0 ~{\rm T}$.
The ion temperature $T_{\rm i} = 1.0 ~{\rm keV}$ and
the density $n_{\rm e} = 0.5 \times10^{19} ~{\rm m^{-3}}$ at the axis.
The equilibrium magnetic field is again constructed by a widely-used
equilibrium code for three-dimensional field, VMEC.
The plasma collisionality and the rotational transform are shown
in Fig.~\ref{fig:param_lhdTe1n05}.

\begin{figure}[t]
  \centering
   \includegraphics[width=0.4\textwidth]{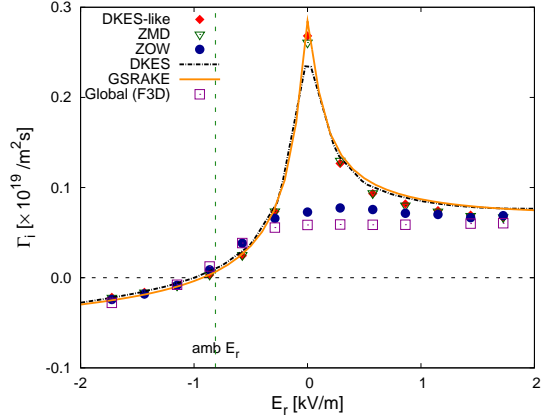}
  \caption{The electric field dependence of $\Gamma_{\rm i}$ at $\sqrt{\psi/\psi_{\rm a}} \simeq 0.29$.
    Following abbreviations are used to represent the particle orbit used in a simulation;
    DKES-like for the particle orbit in the DKES-like limit,
    ZOW for the Zero Orbit Width limit,
    and ZMD for the Zero Magnetic Drift limit.
    Results obtained by using DKES, GSRAKE, and FORTEC-3D codes are also plotted,
    where FORTEC-3D is denoted as Global (F3D).
    The ambipolar $\Er$ estimated by GSRAKE code is also shown by a vertical line.}
  \label{fig:flxEr1}
\end{figure}

\begin{figure}[t]
  \centering
   \includegraphics[width=0.4\textwidth]{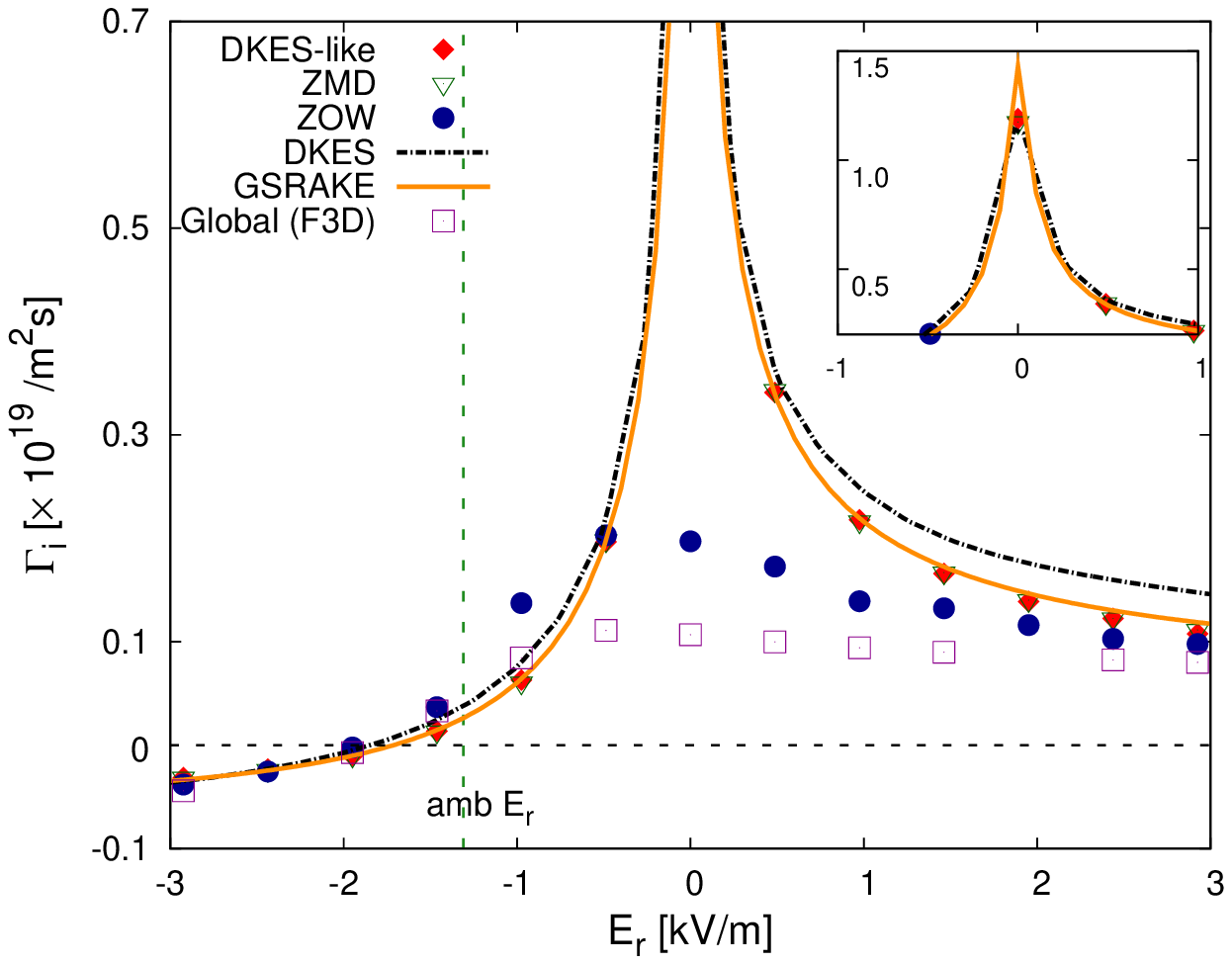}
  \caption{The electric field dependence of $\Gamma_{\rm i}$ at $\sqrt{\psi/\psi_{\rm a}} \simeq 0.49$.
    An Enlarged view for $-1 \le \Er \le 1$ is also shown in the upper right of the figure.
    Legends in the figure are the same as Fig.~\ref{fig:flxEr1}.
    The ambipolar $\Er$ estimated by GSRAKE code is also shown by a vertical line.}
  \label{fig:flxEr2}
\end{figure}

\begin{figure}[t]
  \centering
   \includegraphics[width=0.4\textwidth]{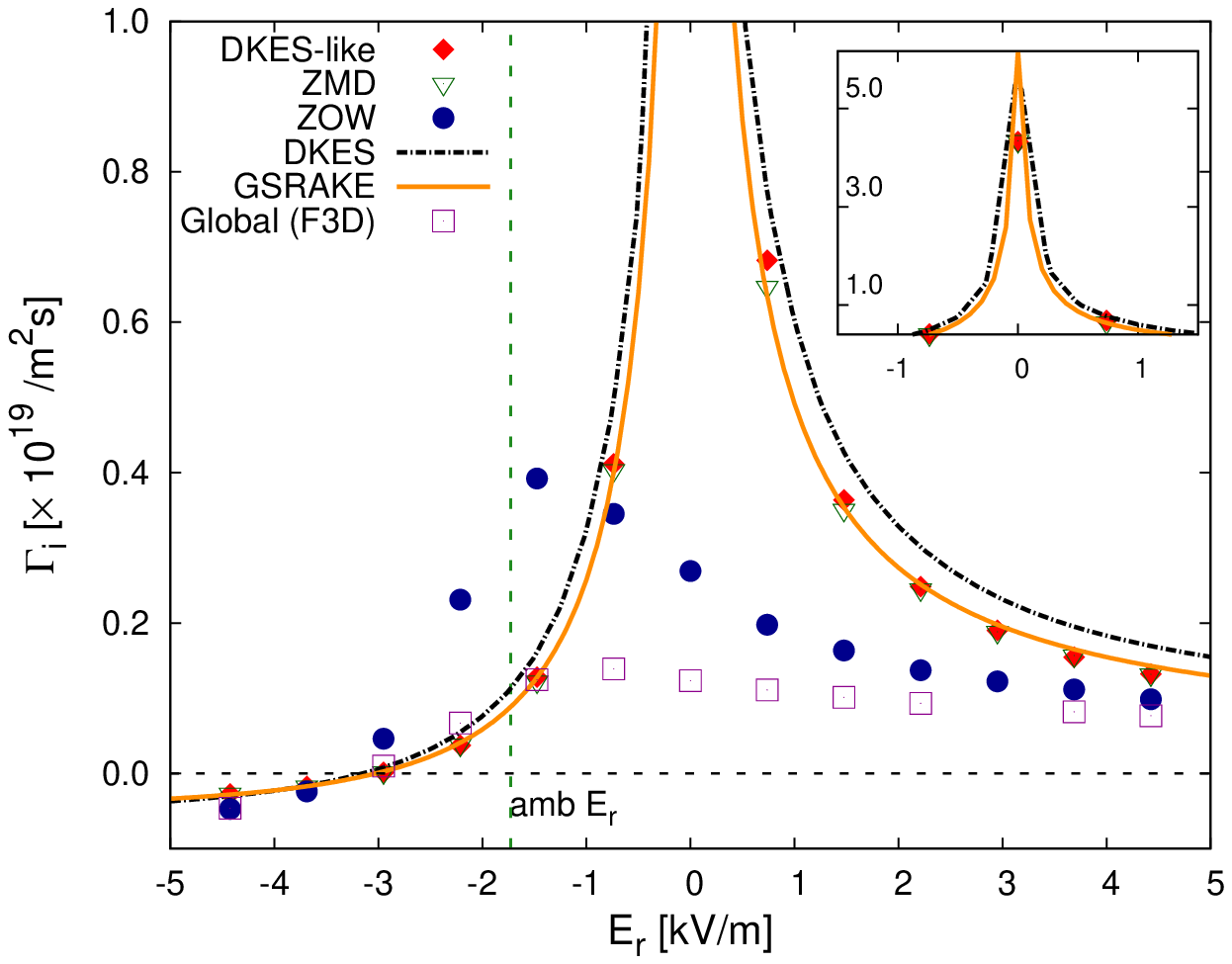}
  \caption{The electric field dependence of $\Gamma_{\rm i}$ at $\sqrt{\psi/\psi_{\rm a}} \simeq 0.74$.
    An Enlarged view for $-1.5 \le \Er \le 1.5 $ is also shown in the upper right of the figure.
    Legends in the figure are the same as Fig.~\ref{fig:flxEr1}.
    The ambipolar $\Er$ estimated by GSRAKE code is also shown by a vertical line.}
  \label{fig:flxEr3}
\end{figure}

The $\Er$ dependence of the neoclassical particle flux
at $\sqrt{\psi/\psi_{\rm a}} \simeq 0.29$, $0.49$ and $0.74$ are shown
in Fig.~\ref{fig:flxEr1} - \ref{fig:flxEr3}.
$\Er$ is given as a constant parameter for each simulation.
The local orbit in ZOW (w/ $\vb$), ZMD (w/o $\vb$) and DKES-like limits are used
in evaluating the flux by the local code developed in this paper.
The full collision operator including the field-particle one is used for all orbit types.
The particle flux by DKES, GSRAKE and FORTEC-3D (denoted as Global, F3D) are also shown in the figures.
The former two are the local codes, while the latter is the global code.
DKES code used here includes the momentum correction.~\cite{Maaßberg2009}
GSRAKE solves the bounce-averaged drift kinetic equation
with the local DKES-like orbit, and only the pitch angle scattering collision operator
without the momentum correction is used.
As mentioned above, FORTEC-3D code evaluates the neoclassical transport with the FOW effect and $\vbh$.

From Fig.~\ref{fig:flxEr1} - \ref{fig:flxEr3},
the particle flux of DKES-like and ZMD limits reproduce
almost the same $\Er$ dependence as GSRAKE at every magnetic surface.
Only a slight difference from original DKES code is also observed.
The momentum conservation does not affect the local neoclassical transport
due to the low collisionality of the plasma considered here.
Also, the results of the local code in DKES-like and ZMD limits well agree with each other
except for a very small difference in small $\Er$ of $\Er \simeq 1.0 ~{\rm kV/m}$,
indicating that the mono-energetic particles and incompressible $\exb$ drift
does not affect so much on the resultant neoclassical transport.
The insignificance of these two assumptions are accounted for as follows.
As Landreman {\it et al.} pointed out,
the mono-energetic assumption varies
the fraction of trapped- and untrapped-boundary in the velocity space,
leading to the underestimation of the trapped particle with DKES-like orbit,~\cite{Landreman2011}
and the neoclassical transport in the DKES-like limit
begins to give a different prediction from that in the ZMD limit
when the poloidal mach number $M_{\rm p}$ exceeds approximately $0.3$.~\cite{Landreman2014}
Since $\Er$ considered here is small enough to satisfy the drift ordering
of $v_{\rm E}/v_{\rm th} \sim \mathcal{O}(\delta)$,
the difference in the trapped-particle fraction
in DKES-like orbit (mono-energetic particles) and ZMD orbit (energy-distributed particles)
does not appear so much.

The results of ZOW limit and global code (FORTEC-3D) clearly show
a different dependence of $\Gamma_{\rm i}$ on $\Er$
from that of other local limits and codes.
The particle flux of other local simulations show a peak at $\Er = 0$ at every surface.
The large $\Gamma_{\rm i}$ there is caused by
the helically-trapped particles which have a large step size in the radial direction.
This can be understood from the discussion around eq.(9)
in Park \textit{et al}.~\cite{Park2009}
Consider a case of zero tangential magnetic drift
in the radially local system (i.e. $\omega_B = 0$).
All the helically-trapped particles cannot move along the surface and remain trapped
when $\Er = 0$ ($\omega_{E} = 0$), giving rise to the large radial transport.
This is the reason why the radial flux in conventional local codes
is enhanced and shows a strong peak  at $\Er = 0$.
On the other hand, the particle flux with the ZOW orbit has
no clear peak at $\sqrt{\psi/\psi_{\rm a}} \simeq 0.29$ in Fig.~\ref{fig:flxEr1},
and shows small peaks at the small negative $\Er$ at $\sqrt{\psi/\psi_{\rm a}} \simeq 0.49$ and $0.74$
in Figs.~\ref{fig:flxEr2} and \ref{fig:flxEr3}, respectively.
The similar tendency is also seen in the results of the global code.
The existence of $\vbh$ in ZOW limit (and the global code) makes
helically-trapped particles move along the surface even without the $\exb$ drift.
Hence, the so-called poloidal resonance occurs at finite $\Er$
since the resonance condition $\omega_B + \omega_E = 0$ is satisfied with the finite $\Er$.
Again from eq.~(9) of Park \textit{et al.,}~\cite{Park2009}
the magnetic precession frequency $\omega_B$ becomes positive for ion species,
indicating that the resonance occurs when the $\exb$ precession is negative
and the radial transport is enhanced at the negative $\Er$.
Moreover, the fraction of trapped particles which satisfies the resonance condition $\omega_B + \omega_E = 0$
is reduced due to the $v$-dependence of $\omega_B$, compared to the the zero $\vbh$ case.
This makes the peaked radial flux broader and smaller in the ZOW limit.

The decrease of $\Gamma_{\rm i}$ at the resonance also arises due to the tangential drift.
The complicated orbit by the combination of $\vbh$ and $\exb$ drifts
causes a transition from trapped to untrapped particles along the surface.
This is called the collisionless detrapping of the particle,
which leads to almost no clear peak near the axis (Fig.~\ref{fig:flxEr1}),
or small and broad peak around mid-radii (Fig.~\ref{fig:flxEr2} and \ref{fig:flxEr3}).
Since the collisionality is low towards the plasma outer region,
the effect of the orbit arises more significantly at $\sqrt{\psi/\psi_{\rm a}} \simeq 0.74$
than at other two surfaces.
Finally, the results of FORTEC-3D show somewhat smaller $\Gamma_{\rm i}$
at every surface compared to those in ZOW limit,
while the peak positions of these two cases are almost the same
($\Er \simeq -1.6 ~{\rm kV/m}$ at $\sqrt{\psi/\psi_{\rm a}} \simeq 0.74$).
This is explained as follows.
The peak position is determined by the balance between $\vbh$ and $\exb$ drift,
and it agrees with each other
since the same tangential magnetic drift is used in the ZOW limit and the global code.
On the other hand,
the finite radial drift is the only included in the global code.
The additional effect
causes another collisionless detrapping, leading to smaller $\Gamma_{\rm i}$.
Also, the extent to which $\Gamma_{\rm i}$ shows smaller value than that of the ZOW case
becomes larger towards the plasma edge in Figs.~\ref{fig:flxEr1} - \ref{fig:flxEr3}.
This is attributed to the larger variation of the magnetic field experienced by the particle along the radial drift,
$\dot{\psi}\partial B/\partial \psi$ towards the edge in the LHD configuration,
resulting in the larger fraction of the detrapped particles and smaller $\Gamma_{\rm i}$ of the global code
than those in the ZOW limit.

\section{Summary}
\label{sec:summary}
In this paper, we provide an alternative way for numerical evaluation of the local neoclassical transport
with several kinds of the particle orbit.
It aims to develop a new local neoclassical transport code which includes
the finite magnetic drift tangential to a flux surface.
Such particle orbit is called the zero orbit width (ZOW) limit in this paper
since the radial drift is only ignored in the drift kinetic equation.
The drift kinetic equation and its variations with various local assumptions
are systematically derived from the global version with the finite orbit width (FOW) effect
to the ZOW, zero magnetic drift (ZMD) and DKES-like limits.
The systematic derivation enables us to investigate the effect of the tangential magnetic drift
in the local neoclassical transport by comparing to the global neoclassical transport
and other local transport models.
The most significant change in the ZOW limit
is that the finite tangential magnetic drift gives rise to the compressibility of the phase-space
volume in the radially local (four-dimensional) phase space.
The conservative property of the phase-space volume,
which varies depending on the orbit and phase space considered,
is discussed in detail.
Based on the discussion of Hu and Krommes,
such non-Hamiltonian (non-conservative) system can be appropriately treated
by regarding the compressibility as a source term to the system.
With this formulation,
the two-weight $\delta f$ Monte Carlo method is presented.

It is worth describing the difference between our source term and that proposed in Landreman {\it et al}.~\cite{Landreman2014}
As discussed in the reference,
the {\it full} and {\it partial} trajectories give rise to a singular perturbation problem
in surface averaged conservation laws of the particle number and energy when the $\Er$ approaches to $0$,
where the full trajectory corresponds to the ZMD orbit in this paper.
This is due to the fact that only the $\Er$ term survives
in the conservation equations for the full and partial trajectory models,
leading to an unphysical behavior of the distribution function in the $\Er = 0$ limit,
see eqs.~(23) and (26) in the reference.
The singular perturbation problem is successfully eliminated
by introducing particle and/or heat source terms in the drift kinetic equation.
On the other hand, in this paper,
the compressibility of the phase-space volume arising from the finite tangential magnetic drift
acts as a source term, which is of a higher order in the drift kinetic equation in the ZOW limit.
Since the compressibility changes the conservation equations, 
the singular perturbation problem is avoided. 
It should be emphasized that
our source term is introduced to practically evaluate the neoclassical transport in the ZOW model
by solving the non-Hamiltonian drift kinetic equation as an initial value problem.
Although the source term actually violates the particle number and/or energy conservations,
in most cases presented here except for the near-axis region,
it does not cause any matter in evaluating neoclassical particle and energy fluxes.
The impact of the source term on the neoclassical transport will be discussed more in detail in future works.

The code verification and its validity are checked by theoretical and numerical benchmark
calculations for axisymmetric and non-axisymmetric magnetic field configurations.
For a tokamak case, the neoclassical ion thermal diffusivity in the axisymmetric plasma well reproduces the Chang-Hinton formula
in a wide range of the collisionality.
Also, the parallel flow coefficient of the local code
with DKES-like orbit and the ZOW orbit
are shown to well agree with the theoretical estimations of Hirshman-Sigmar.
This indicates that the finite magnetic drift does not change
the conventional local neoclassical transport so much in an axisymmetric configuration.

In a non-axisymmetric device, the finite tangential magnetic drift significantly changes
the local neoclassical transport.
In conventional local neoclassical transport calculations (ZMD and DKES-like limits),
the poloidal resonance condition, $\hat{v}_{\rm B} \simeq v_{\rm E}$,
where the helically-trapped particle causes a large radial transport,
is satisfied with $\Er = 0$ due to the absence of $\vbh$.
When $\vbh$ exists (in the ZOW limit), however,
the poloidal resonance is shifted to a small negative $\Er$.
It is demonstrated for the first time that the finite tangential magnetic drift
gives rise to a qualitative change
in the dependence of the neoclassical transport on the radial electric field
even in the local neoclassical transport model.
Also, similarly to the global neoclassical transport,
the large radial transport at the resonance seen in the ZMD and DKES-like limits
can be avoided in the ZOW limit due to the collision detrapping along the local orbit.
Hence, two important physics included in the global code can be captured by the local code in the ZOW limit.

A key role of the neoclassical transport in a non-axisymmetric plasma
is to predict the ambipolar $\Er$ according to the ambipolar condition of the neoclassical particle flux.
As demonstrated in the paper,
the finite magnetic drift changes the dependence of the ion particle flux on $\Er$.
The ambipolar $\Er$ predicted can vary depending on whether $\vbh$ is included
in evaluating the neoclassical transport.
The main cause of the difference comes from the shift of the poloidal resonance condition,
and it is included in our local code in the ZOW limit.
Since the local code is less time-consuming and requires less computational cost
than the global code,
the local code will be a preferable alternative to predict the ambipolar $\Er$
in experimental analyses.

\begin{acknowledgments}
  The authors would like to acknowledge Dr. J.~L.~Velasco 
  for kindly providing numerical results by DKES code
  and to thank Dr. J.~M.~Garc{\'i}a-Rega{\~n}a 
  for useful information on EUTERPE code.
  This work was carried out using the HELIOS supercomputer system
  at Computational Simulation Centre of International Fusion Energy Research
  Centre (IFERC-CSC),
  Aomori, Japan, under the Broader Approach collaboration between Euratom and Japan,
  implemented by Fusion for Energy and JAEA.
  This work was supported in part by JSPS Grant-in-Aid for Young Scientists (B), No. 23760810,
  NIFS Collaborative Research Programs NIFS13KNST051, NIFS13KNST060, and NIFS14KNTT026.
\end{acknowledgments}

\bibliography{localNCT_150217}

\end{document}